# *p*-wave magnet and hedgehog-type Berry curvature in helimagnetic $MnAu_2$

Hogyun Jeong[1, †], Sangyun Lee[1, †], Junhee Shin[1], Jung-Woo Yoo[2], Changhee Sohn[1], Yoon Seok Oh[1] & Hosub Jin[1, *]

[1]*Department of Physics, Ulsan National Institute of Science and Technology, Ulsan 44919, Korea*

[2]*Department of Materials and Science and Engineering, Ulsan National Institute of Science and Technology, Ulsan 44919, Korea*

**Abstract**

Recently discovered altermagnetism in collinear compensated magnets shows even-parity spin texture in momentum space. Beyond the collinear spin ordering, unique odd-parity spin textures emerge in noncollinear compensated magnets. The noncollinear candidates, however, remain unexplored toward a room-temperature metallic altermagnet. Here, we demonstrate that $MnAu_2$ exhibits metallic *p*-wave magnetism and large spin splitting induced by helical spin ordering. By adapting the band-unfolding scheme based on the translation operators combined with spin rotations, first-principles calculations reveal an unconventional Fermi surface around the $\widetilde{M}$-point composed of a single electron pocket with *p*-wave spin texture. Moreover, the spin twist in the helimagnet triggers topologically non-trivial hedgehog Berry curvature, which links to the nonlinear Hall effect and spin Hall effect. Considering the experimental $T_C$ = 335 to 370 K, $MnAu_2$ could establish itself as an ideal candidate for a room-temperature metallic *p*-wave magnet, promising for versatile spintronic applications.

† These authors contributed equally to this work

* Corresponding author email: hsjin@unist.ac.kr

*Introduction*—As a new member of collinear compensated magnets, altermagnetism has been proposed and subsequently discovered by revealing a unique spin texture in reciprocal space [1-12]. In altermagnetic systems, the time-reversal and rotational symmetry combine to connect the compensated pairs in each magnetic sublattice. This compensation symmetry is accompanied by specific lattice distortions, leading to anisotropic spin bands and even-parity spin texture [1-8,10-12]. Altermagnetic spin splitting results in various novel phenomena, including spin splitting torque [13,14], tunnelling magnetoresistance [15-17], and efficient magnon transport [18-20]. Currently, *room-temperature metallic altermagnets with large spin splitting* are under intensive investigation for further advancements [21,22].

Extending our scope beyond collinear to noncollinear compensated magnets may offer new opportunities in the realm of altermagnetism applications. In contrast to collinear altermagnets, where lattice distortions drive even-parity spin textures, some noncollinear compensated magnets exhibit odd-parity spin textures [23-26] due to the spatial variation of the spin axis, leading to various potential applications in spintronics. Furthermore, this expansion to noncollinear magnetism increases the possibility for developing room-temperature metallic altermagnets with large spin splitting. However, there is still a lack of exploration into viable noncollinear candidates [25,27-29].

Here, we theoretically demonstrate that helimagnetic $MnAu_2$ is a candidate for a room-temperature metallic *p*-wave magnet. In the continuum and tight-binding models, the twist in spin-space caused by helical magnetic ordering creates an effective gauge field proportional to the *z*-component of spin angular momentum, leading to *p*-wave spin texture. By unfolding the band structure of $MnAu_2$ onto the primitive unit cell, an unconventional single Fermi surface with a *p*-wave spin texture emerges around the $\widetilde{\mathrm{M}}$-point in the unfolded Brillouin zone. Along the $\widetilde{\mathrm{M}}^*$-$\widetilde{\Gamma}$-$\widetilde{\mathrm{M}}$ line, two *p*-wave bands with opposite polarity are fully separated by approximately 1.5 eV, with one of them crossing the Fermi level. In addition, helical spin ordering triggers quantum geometry encoded in the Bloch states, leading to hedgehog-type Berry curvature (BC). This topologically non-trivial hedgehog

BC becomes the source of a large nonlinear Hall effect by inducing a BC dipole. Accompanied by the hedgehog BC and $p$-wave spin texture, a substantial spin Hall effect also arises from the spin BC. With its experimental critical temperature [30-32], $MnAu_2$ could be the best candidate for a room-temperature metallic $p$-wave magnet with large spin splitting and serve as a versatile platform showing efficient nonlinear transverse transport and spin-charge conversion.

*Effective gauge field induced by spin-space twist*—In helimagnets, the $p$-wave spin texture arises solely from twist in spin-space, without any distortion in the lattice. Bloch electrons traversing through the helical spin structure experience their non-inertial reference frame of the rotating spin axes. Depending on the direction of the electron motion, the spin reference frame rotates either clockwise or counterclockwise, generating the spin-dependent effective gauge field (Fig. 1a, b). The effective gauge field is demonstrated in a continuum model of a helimagnetic chain expressed as $\mathcal{H} = \frac{\mathbf{p}^2}{2m} - J\mathbf{S}\cdot\mathbf{M}(z)$, where $\mathbf{S}$ denotes the electron's spin, and $\mathbf{M}(z) = M(\cos qz, \sin qz, 0)$ indicates the continuous helical magnetic moment. By applying a local spin rotation, $\mathcal{C} \equiv \exp\left(-\frac{i}{\hbar} qzS_z\right)$, which maps the curved spin-space onto a flat one, the continuum model becomes [33,34]

$$\tilde{\mathcal{H}} \equiv \mathcal{C}^{\dagger}\mathcal{H}\mathcal{C} = \frac{1}{2m}\left[{p_x}^2 + {p_y}^2 + (p_z - qS_z)^2\right] - JMS_{\tilde{x}} = \frac{\mathbf{p}^2}{2m} - \frac{q}{m}p_z S_z - JMS_{\tilde{x}}, \quad (1)$$

where $S_{\tilde{x}}$ denotes the spin component parallel to the helical moment. In a spin-rotating frame, the effective gauge field $-qS_z$ emerges, which couples electron momentum and spin and is responsible for the $p$-wave spin texture. The third term in Eq. (1) represents the Zeeman-like exchange interaction between conducting electrons and helimagnetic moments. It provides additional energy splitting depending on whether the in-plane spin components of the electron are parallel or antiparallel to $\mathbf{M}(z)$. Still, there is no spin texture induced by the Zeeman-like exchange term due to the net zero moment of the helimagnet (See Supplementary Material for more details of the continuum model) [35].

We show that the simplest $s$-orbital tight-binding model of a helimagnetic chain, where the local moments rotate by $2\pi/3$ between adjacent sites, hosts *p*-wave magnetism. The tight-binding Hamiltonian is written as follows

$$\mathcal{H}_{\mathrm{TB}}(k_z) = \begin{pmatrix} f_0 & h^\dagger(k_z) & h(k_z) \\ h(k_z) & f_1 & h^\dagger(k_z) \\ h^\dagger(k_z) & h(k_z) & f_2 \end{pmatrix}, \tag{2}$$

where $f_n = -\frac{\hbar}{2} JM \begin{pmatrix} 0 & e^{-i\frac{2\pi}{3}n} \\ e^{i\frac{2\pi}{3}n} & 0 \end{pmatrix}$ $(n = 0,1,2)$ and $h(k_z) = -te^{-\frac{ik_z c}{3}} \mathbb{I}_{2\mathrm{x}2}$ . Here, the on-site energy of the *s*-orbital is set to be zero, $t$ is the nearest neighbor hopping amplitude, and $c$ denotes the lattice constant containing the helimagnetic superstructure (Fig. 1c). The electronic structure and spin texture of the tight-binding model are shown in Fig. 1d, where the bands clearly exhibit *p*-wave spin texture, but are intricately entangled. The band dispersion shown in Fig. 1d suggests that the spin splitting does not align with the continuum model described in Eq. (1). Consequently, it is difficult to interpret spin splitting in terms of the effective gauge field and Zeeman-like exchange interaction. The representative feature of *p*-wave magnet can be revealed by adapting the generalized band unfolding scheme to the primitive unit cell, where the translation operator combines with spin rotation (see Supplementary Material for more details of the generalized band unfolding scheme) [35]. A Brillouin zone is extended resulting from a change of lattice constant $\left(\tilde{c} = \frac{1}{3}c\right)$ and a momentum shift of $\frac{\pi}{c}$ is induced by the spin-rotation, leading two fully separated bands with opposite polarities (Fig. 1e). At $\tilde{k}_z = 0$ in the unfolded Brillouin zone, the Bloch states exhibit a null spin expectation value, and the energy splitting between the two bands depends solely on whether the electron spins are parallel or antiparallel to the helimagnet. As the manifestation of the $S_z$-dependent effective gauge field induced by spin twisting, the traversing Bloch state with finite $\tilde{k}_z$ gains extra spin polarization parallel to the helical axis, resulting in the *p*-wave magnet.

*Helimagnetic MnAu2 as a candidate for p-wave magnetism*—Here, we examined the electronic structure of helimagnetic $MnAu_2$ as a potential candidate for *p*-wave magnetism. As shown in Fig. 2a, $MnAu_2$ has a body-centered tetragonal structure belonging to the space group *I4/mmm* [40]. In a helimagnetic state, each Mn layer in the (001) plane exhibits ferromagnetic ordering in the $ab$-plane, and helical spins with a 45° rotation between adjacent Mn layers emerge (see End Matter for detailed information on the crystal and magnetic structure of $MnAu_2$). To investigate the $\mathbf{q} = \left(0,0,\frac{1}{4}\frac{2\pi}{c}\right)$ helimagnetic ordering, we performed density functional theory calculations within an elongated supercell that contains eight formula units of $MnAu_2$. As shown in Fig. 2b, opposite spin textures in the $k_z = +\frac{\pi}{8c}$ and $k_z = -\frac{\pi}{8c}$ planes, along with null spin expectation values in the $k_z = 0$ plane, demonstrate a $p$-wave magnet. However, these results do not accurately capture the representative band dispersion and spin splitting in the $p$-wave magnet, as illustrated in Fig. 1e. Using the generalized band unfolding scheme, we can verify non-degenerate bands with $p$-wave spin texture in the unfolded Brillouin zone (Fig. 2c, d). The primitive cell that defines the unfolded Brillouin zone is constructed using the unit vectors $\tilde{\boldsymbol{a}}_i$ $(i = 1,2,3)$, where the translation by $\tilde{\boldsymbol{a}}_i$, combined with $\pi/4$ spin rotation, provides the effective translation symmetry. As shown in Fig. 2c, there exists a clear opposite spin texture along the $\tilde{\mathrm{N}}^*$-$\tilde{\Gamma}$-$\tilde{\mathrm{N}}$ line, while there is no spin texture along the $\tilde{\mathrm{X}}$-$\tilde{\Gamma}$ line. This is more evident in the Fermi surface contour plots depicted on various planes of the reciprocal space (Fig. 2d). The Fermi surface consists of several non-degenerate pockets, sheets, and torus, all exhibiting $p$-wave spin textures. In the two horizontal planes ($\tilde{k}_z = \pm\frac{1}{4}\frac{2\pi}{c}$), the Fermi contours show opposite spin textures, while the $\tilde{k}_z = 0$ plane has no spin texture. Notably, the circular electron pocket centered at the $\tilde{\mathrm{M}}$-point on the vertical cut plane is composed of a single band and presents an antisymmetric spin texture, with a spin-zero line at the equator. These features differ from collinear altermagnetism, where spin splitting begins to develop from the spin-degenerate nodal lines [1-8,10-12].

*Characteristics of the Bloch state and spin splitting*—The $\tilde{\mathrm{M}}^*$-$\tilde{\Gamma}$-$\tilde{\mathrm{M}}$ line has the identical symmetry as a one-dimensional helimagnetic chain, and the electronic band structure thus reveals the representative features of the helimagnet-hosted $p$-wave magnet in terms of spin texture and band splitting. As shown in Fig. 3a, two non-degenerate bands split by approximately 1.5 eV emerge near the Fermi level. Among them, only one band crosses the Fermi level, which corresponds to the electron pocket at the $\tilde{\mathrm{M}}$-point in Fig. 2d. These spin-split-off bands demonstrate the antisymmetric $S_z$-spin texture with opposite polarity. At the $\tilde{\Gamma}$-point, the real-space spin texture of the lower (upper) band is aligned parallel (antiparallel) to the local magnetic moment of Mn (Fig. 3b). Therefore, spin splitting at the $\tilde{\Gamma}$-point originates from a Zeeman-like exchange term, and helical spin texture ensures zero spin expectation value. The traveling Bloch states with a finite $\tilde{k}_z$ acquire an additional $S_z$ component, exhibiting conical spin texture (Fig. 3c, d). The Bloch state displays a helical texture induced by the Mn moment and an $S_z$ texture caused by an effective gauge field at the same time. The real-space spin textures of two split-off bands at the same $\tilde{k}_z$ are opposite, and only the $S_z$ component is opposite for the $\pm\tilde{k}_z$ partner. The real-space spin texture of the Bloch state in each band shows that the effective gauge field and Zeeman-like terms in the continuum model are the main sources of the $p$-wave spin texture and spin splitting in momentum space.

*Topologically non-trivial hedgehog Berry curvature*—Under symmetry-broken environments, quantum geometry encoded in the wavefunction appears, inducing various nonlinear anomalous transports [41-43]. The spin-space twist of the helimagnetic $MnAu_2$ not only gives rise to $p$-wave magnetism but also hosts a hedgehog BC texture. In the electronic structure calculations that include the spin-orbit coupling, BC vectors near the Z-point fan out radially in the $k_x$-$k_y$ plane and point inward along the $k_z$ line (Fig. 4a). As shown in Fig. 4b, reversing the handedness of magnetic ordering switches BC texture from hedgehog to anti-hedgehog, demonstrating that the helical spin ordering is a source of the topologically nontrivial BC distribution. These unique textures produce a switchable BC dipole measurable in nonlinear Hall experiments (Fig. 4c). Its magnitude reaches 0.1 at the Fermi level, which is comparable to the significantly enhanced value at the topological phase transition in pressurized BiTeI [44]. In addition to the $p$-wave spin texture and hedgehog BC texture, spin BC arises, resulting in the spin Hall effect. As shown in Fig. 4d, $MnAu_2$ exhibits large spin Hall conductivity close to the values of Au and Ir [45,46]. Therefore, the helimagnetic $MnAu_2$ can serve as a versatile platform for spintronics, facilitating efficient spin generation, spin-charge conversion, and nonlinear anomalous charge transport.

As a remark, we included the spin-orbit coupling for calculating BC and spin BC. The spin-orbit coupling has a minimal impact on both the electronic structure near the Fermi level and the *p*-wave spin texture. Additionally, the value of $\mathbf{q} = \left(0, 0, \frac{1}{4}\frac{2\pi}{c}\right)$ used in our calculations differs slightly from the values reported in the experiment, which vary from $\left(0, 0, \frac{1}{3}\frac{2\pi}{c}\right)$ at 5 K to $\left(0, 0, 0.22\frac{2\pi}{c}\right)$ at 250 K [31,47]. To see the effect of the difference in $\mathbf{q}$ value on the electronic structure, we applied the generalized Bloch theorem in our first-principles calculations. Our results confirm that the *p*-wave spin texture remains intact for the theoretically optimized value of $\mathbf{q} = \left(0,0,0.277\frac{2\pi}{c}\right)$[35,39].

Previous experimental measurements on the helimagnetic ordering of $MnAu_2$ also suggest a potential occurrence of room-temperature metallic *p*-wave magnetism. The reported critical temperature is 335-370K [30-32]. In particular, the robustness of spin-spiral helimagnetic ordering at room temperature is evidenced by the fact that the measured local magnetic moment retains 2.82 $\mu_B$ at 300K which is about 80% of its 10K value reported as 3.50 $\mu_B$ [31].

*Conclusion*—A moving electron in a noncollinear magnetic system acquires its local spin axes as a non-inertial reference frame in spin space, resulting in an effective field that depends on both spin and momentum. It enables non-relativistic spin splitting accompanied by an odd-parity spin texture. Our calculations confirmed that the helimagnetic compound $MnAu_2$ exhibits a non-relativistic *p*-wave spin texture near the Fermi surface. Combined with the experimental $T_C$ above 300 K, this suggests that $MnAu_2$ is a potential candidate for room-temperature metallic altermagnetism with large spin splitting. Moreover, the spin twist in the helimagnet results in the onset of topologically non-trivial hedgehog Berry curvature, leading to the nonlinear Hall effect. Experimental verification of the electronics structure and its validation for spintronic applications are anticipated.

**Acknowledgements** This work was supported by the National Research Foundation of Korea (NRF) funded by the Korea government (MIST) under Grant No. RS-2024-00487988, RS-2023-NR076899, RS-2021-NR057351, and RS-2023-00257666.

# End Matter

*Computational details*—First-principles calculations are done via a code of density-functional theory with Full-potential Linearised Augmented Plane Wave method as implemented in the Elk code [48]. Muffin-Tin radii $r_{\mathrm{MT}}$ of Mn and Au are 2.40 and 2.62 Bohr radius, respectively. $r_{\mathrm{MT}} * |\mathbf{G}+\mathbf{k}|_{\max}$ is set to be 8. Energy linearization is executed by adding local orbitals to Muffin-Tin spheres [49]. The PBE parameterization of generalized-gradient approximation to exchange-correlation functionals is adopted [50]. The LDA+$U$ method is introduced to consider the effect of electronic correlation in magnetic transition metals, and it stabilizes the helimagnetic configurations relative to the ferromagnetic configuration [39]. $U = 5$ eV and $J = 1$ eV are applied to Mn's *d* orbitals with double-counting correction of the fully localized limit [51]. A *k*-point grid of $8 \times 8 \times 3$ is used.

$MnAu_2$ crystal has space group *I4/mmm* (No. 139) [40]. Its Bravais lattice, called body-centred tetragonal, are determined by three primitive vectors $\tilde{\boldsymbol{a}}_1 = (a/2, a/2, -c/2)$, $\tilde{\boldsymbol{a}}_2 = (-a/2, a/2, c/2)$, $\tilde{\boldsymbol{a}}_3 = (a/2, -a/2, c/2)$, where experimental lattice constants are $a = 3.37$ Å and $c = 8.75$ Å. The Wigner-Seitz cell contains one Mn atom and two Au atoms, and their positions in Cartesian coordinates are Mn $(0,0,0)$, Au1 $(0,0, c/3)$, Au2 $(0,0, 2c/3)$.

In experiments, $MnAu_2$ shows its helimagnetic spin ordering with a 60° angle of rotation at 5 K and 40° at 250 K [32,47]. We prefer the (commensurate) supercell method over the generalized Bloch theorem [52] to analyse the helimagnet, as this approach allows for the incorporation of spin-orbit coupling. The magnetic supercell is constructed by repeating the conventional tetragonal 4 times along the *z* direction. The supercell has 8 Mn layers separated by $c/2$ along the *z* direction and Mn's spin moments between adjacent layers make a 45° angle of rotation around the *z* direction, i.e. $\mathbf{q} = \left(0, 0, \frac{1}{4}\frac{2\pi}{c}\right)$.

The intrinsic contribution to the anomalous Hall effect can be explained using the concept of BC

[53]. The linear-response expression for BC is [54]

$$\Omega_{n\mathbf{k}}^{\gamma} = i \sum_{m(\neq n)} \epsilon_{\alpha\beta\gamma} \frac{\langle \psi_{n\mathbf{k}} | v_{\alpha} | \psi_{m\mathbf{k}} \rangle \langle \psi_{m\mathbf{k}} | v_{\beta} | \psi_{n\mathbf{k}} \rangle}{(\omega_{n\mathbf{k}} - \omega_{m\mathbf{k}})^2 + \eta}.$$

Here, $\psi_{n\mathbf{k}}$ means a Kohn-Sham eigenstate with the band index $n$ at $\mathbf{k}$, $\varepsilon_{n\mathbf{k}} = \hbar\omega_{n\mathbf{k}}$ an eigenvalue, $v_{\alpha} = i/\hbar[\mathcal{H}, r_{\alpha}]$ the velocity operator and its Heisenberg equation of motion, Greek letters imply cartesian coordinates, and a positive infinitesimal $\eta$ is introduced to increase numerical stability. $\epsilon_{\alpha\beta\gamma}$ means Levi-Civita symbol with implicit summation over repeated indices. A dipole component of the BC distribution is written as [41,55]

$$D_{\alpha\beta} = \int_{\mathrm{BZ}} \frac{d\mathbf{k}}{(2\pi)^3} \sum_{n} f_{n\mathbf{k}} \frac{\partial \Omega_{n\mathbf{k}}^{\beta}}{\partial k_{\alpha}},$$

where $f_{n\mathbf{k}}$ is the Fermi-Dirac distribution function. The formula for spin Hall conductivity is constructed by replacing one of charge-current operators $-ev_{\alpha}$ with a spin current $\frac{1}{2}\left\{v_{\alpha}, \frac{\hbar}{2}\sigma_{\gamma}\right\}$ in the BC [56],

$$\sigma_{\alpha\beta}^{\gamma} = \frac{e^2}{\hbar} \frac{\hbar}{e} \int_{\mathrm{BZ}} \frac{d\mathbf{k}}{(2\pi)^3} \sum_{n} f_{n\mathbf{k}} \sum_{m(\neq n)} \frac{-\mathrm{Im} \left\langle \psi_{n\mathbf{k}} \middle| \frac{1}{2}\{v_{\alpha}, \sigma_{\gamma}\} \middle| \psi_{m\mathbf{k}} \right\rangle \langle \psi_{m\mathbf{k}} | v_{\beta} | \psi_{n\mathbf{k}} \rangle}{(\omega_{n\mathbf{k}} - \omega_{m\mathbf{k}})^2 + \eta}.$$

Here, $\sigma_{\gamma}$ is the Pauli matrix and the adiabatic spin-current expression is used.

In transport calculations of the helimagnetic supercell, we use a $k$-point grid of $50 \times 50 \times 25$ and include approximately 500 unoccupied states up to 33 eV from the Fermi level. We set $\eta = 10^{-6}$ to smooth BC values because they are too spiky due to heavily folded band-structure. $f_{n\mathbf{k}}$ is given by the Fermi-Dirac distribution at 63 K. Spin-orbit coupling is included for transport calculations.

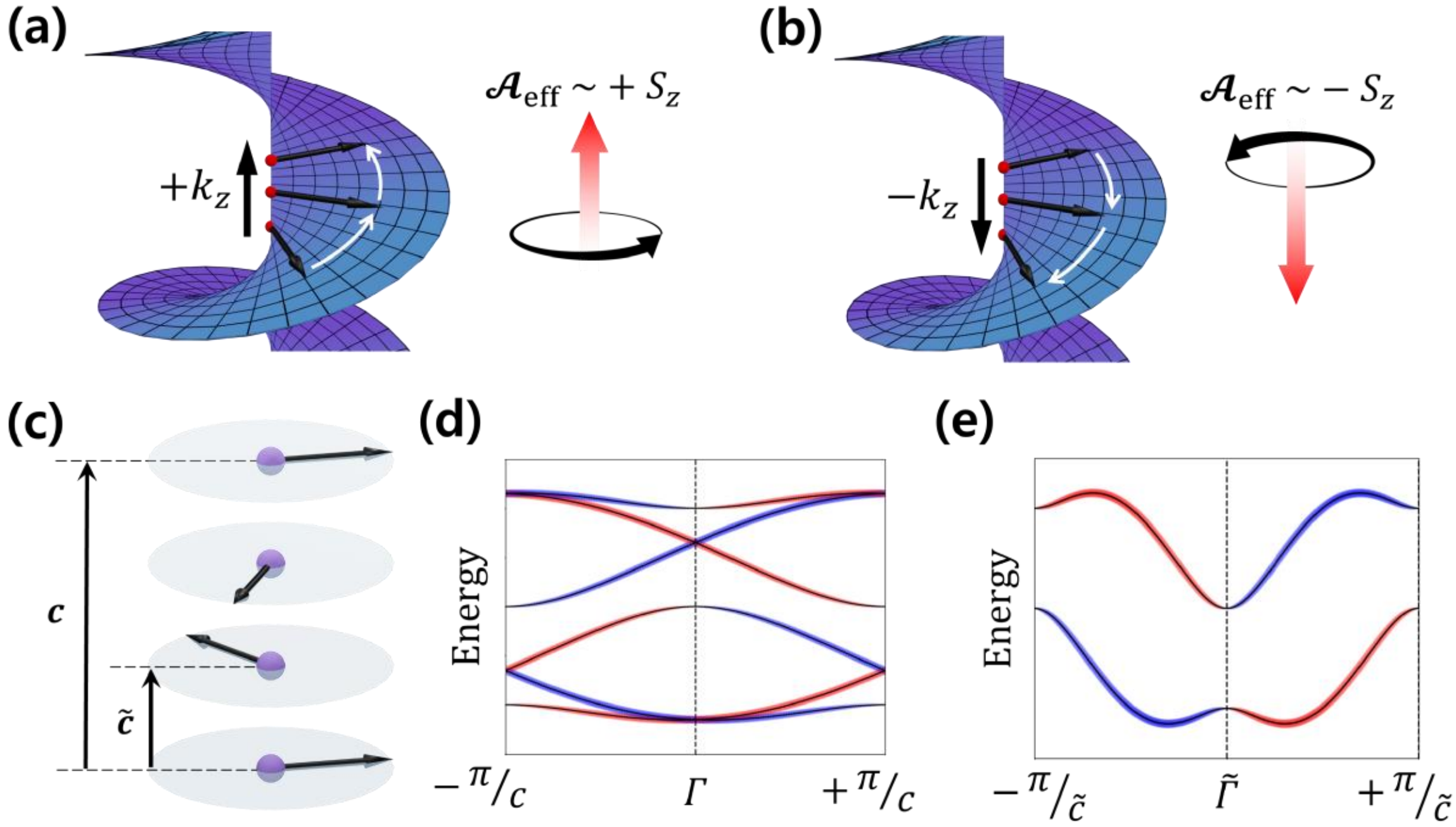

Figure 1. Origin of $p$-wave magnetism in helimagnets. (a, b) Schematic illustration of an effective gauge field $\mathcal{A}_{\text{eff}}$ arising from a twisted spin space. The electron movement along $+z$ ($-z$) direction generates a counterclockwise (clockwise) rotation of its spin coordinate, respectively. (c) The crystal and magnetic structure of a threefold helimagnetic chain is depicted, where $c$ and $\tilde{c}$ indicate the lattice constants of the magnetic supercell and primitive cell, respectively. The rotating black arrows indicate magnetic moments. The band structures with $\langle S_z \rangle$ texture are shown within the folded (d) and unfolded (e) Brillouin zones. The line width denotes the magnitude of the $S_z$ expectation value. Red (blue) corresponds to positive (negative) values.

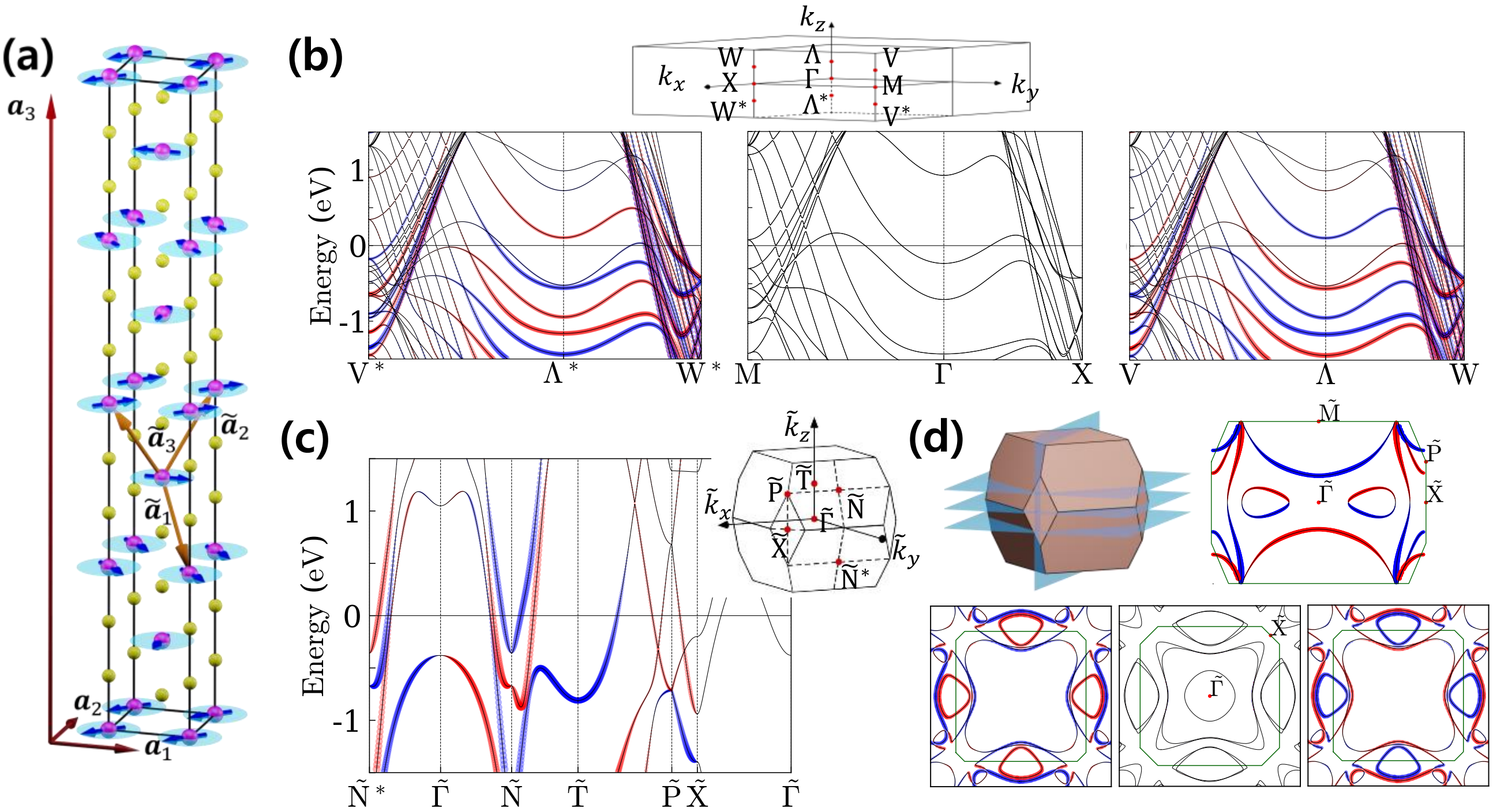

Figure 2. Metallic $p$-wave magnetism in helimagnetic $MnAu_2$. (a) Crystal and magnetic structures of helimagnetic $MnAu_2$. The purple and yellow spheres indicate Mn and Au atoms, respectively. The blue arrows attached to Mn atoms represent magnetic moments. $\boldsymbol{a}_i$ and $\tilde{\boldsymbol{a}}_i$ denote the unit vectors of the supercell and primitive cell, respectively. (b) The band structure with the $S_z$ expectation value is presented within the Brillouin zone of the magnetic supercell. The top panel depicts the Brillouin zone and the $k_z$ axis is scaled by a factor of three. (c) The unfolded band structure with $\langle S_z \rangle$ value is plotted. Band unfolding involves an additional momentum shift of $\frac{\mathbf{q}}{2}$ resulting from the spin rotation. The top-right inset shows the unfolded Brillouin zone. Here, the overhead tilde indicates the $k$-points in the unfolded Brillouin zone, and an asterisk denotes their inversion partners. (d) Fermi surface contours with $\langle S_z \rangle$ spin textures on the selected planes in the Brillouin zone of the primitive cell. The top-left panel shows the selected planes. The top-right panel corresponds to the (-110) vertical plane, and the bottom panels are arranged in order of $\tilde{k}_z = -\frac{2\pi}{4c}, 0, \frac{2\pi}{4c}$ planes. Colour schemes in (b)-(d) are identical to those in Figure 1.

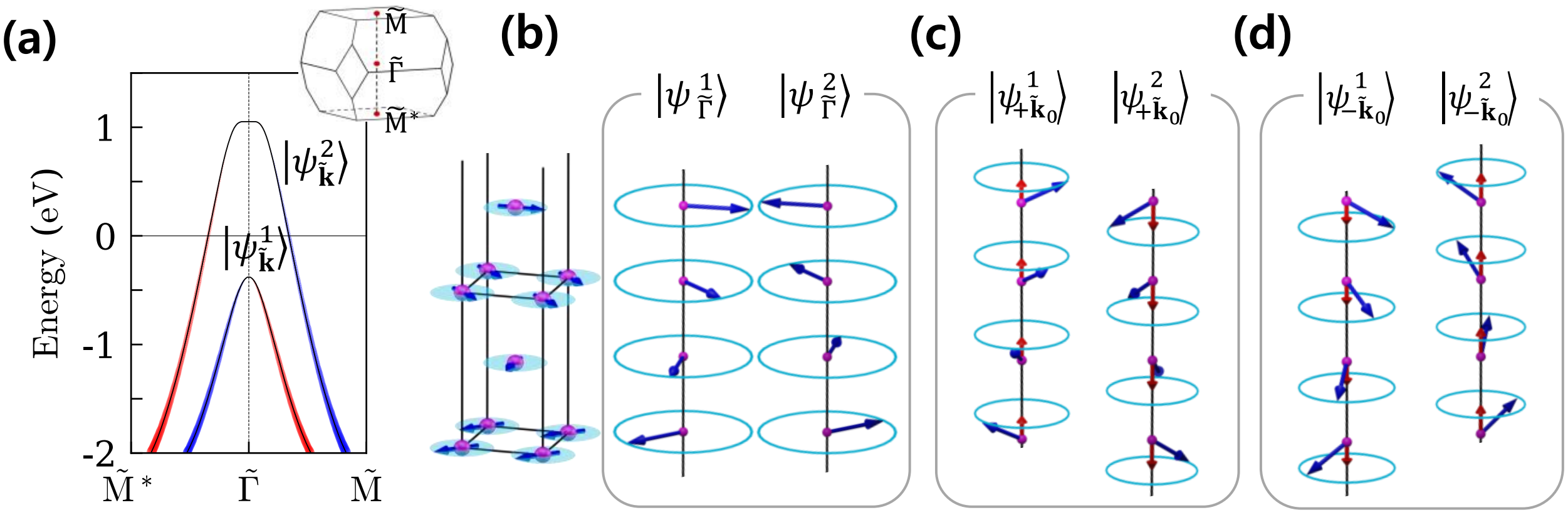

Figure 3. Real-space spin texture of the Bloch states. (a) The band structure with $\langle S_z \rangle$ spin texture is plotted along the $\widetilde{\mathrm{M}}^*$-$\tilde{\Gamma}$-$\widetilde{\mathrm{M}}$ line of the unfolded Brillouin zone, as shown in the inset. (b-d) Real-space spin textures of the two spin-split-off Bloch states are presented. The blue arrows indicate the spin expectation value projected onto Mn atoms, and the red arrows denote the *z*-component spin moment. Here, $\tilde{\mathbf{k}}_0 = \frac{3}{8}\widetilde{\Gamma\mathrm{M}}$. The left figure in (b) denotes the corresponding Mn atoms with helimagnetic ordering.

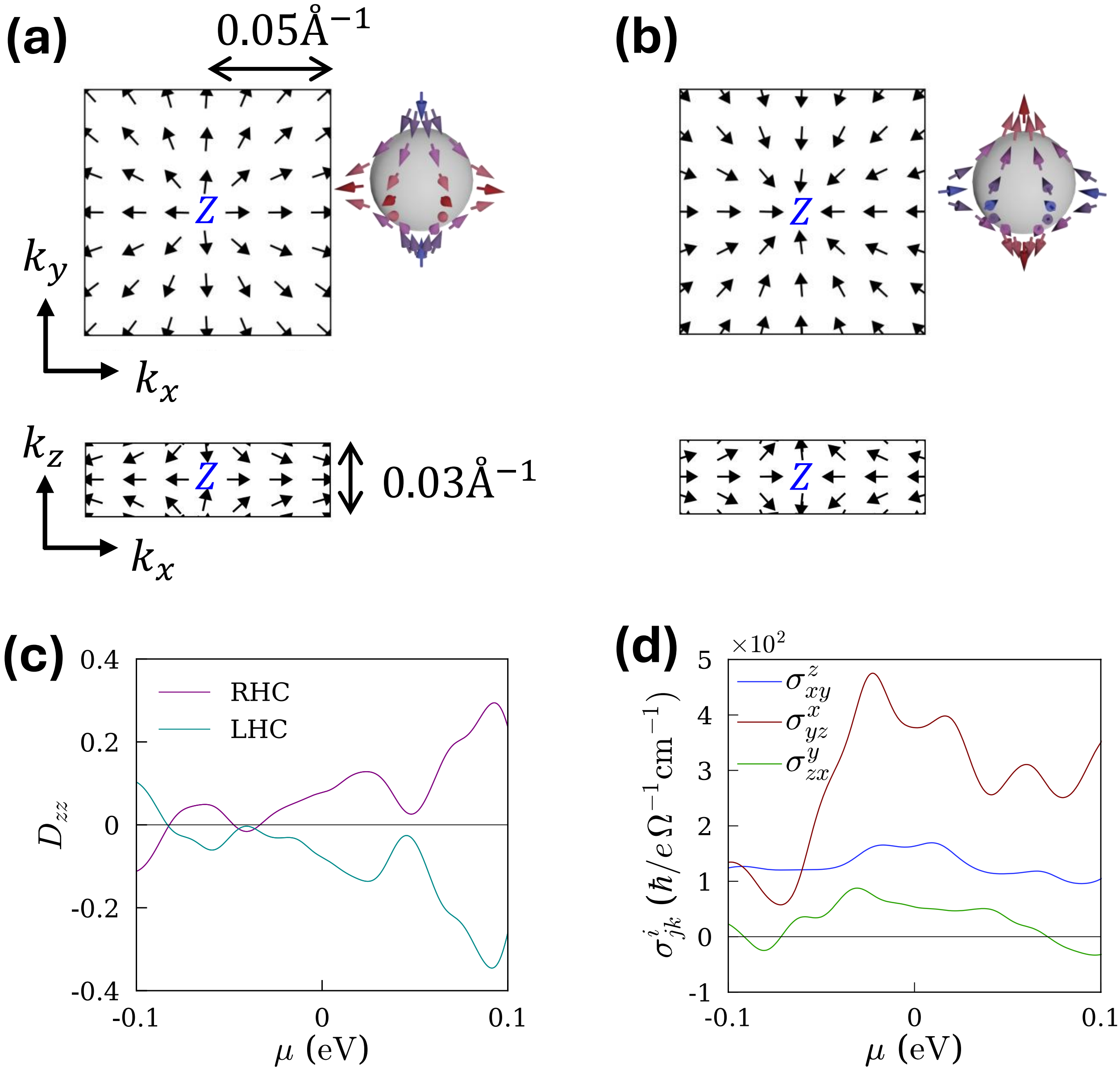

Figure 4. Topologically non-trivial hedgehog Berry curvature textures. (a) and (b) represent hedgehog-type and anti-hedgehog Berry curvature distributions near $\mathbf{k} = \left(0,0,\frac{1}{2}\frac{2\pi}{4c}\right)$ point (Z point) for right-handed chirality (RHC) and left-handed chirality (LHC), respectively. The black arrows indicate the normalized Berry curvature. (c) Berry curvature dipole ($D_{zz}$) is plotted as a function of the chemical potential ($\mu$). (d) Spin Hall conductivity is shown as a function of $\mu$.